\documentclass[fleqn,10pt]{wlscirep}
\usepackage[utf8]{inputenc}
\usepackage[T1]{fontenc}

\usepackage{pifont}
\newcommand{\cmark}{\ding{51}}%
\newcommand{\xmark}{\ding{55}}%
\usepackage{nameref}
\usepackage{arydshln}
\usepackage{multirow}
\usepackage{float}
\usepackage{colortbl, xcolor}
\usepackage[section]{placeins}

\title{An Evaluation Tool for Backbone Extraction Techniques in Weighted Complex Networks}

\author[1,*]{Ali Yassin}
\author[2]{Abbas Haidar}
\author[3]{Hocine Cherifi}
\author[4]{Hamida Seba}
\author[1]{Olivier Togni}
\affil[1]{University of Burgundy, Laboratoire d’Informatique de Bourgogne, Dijon, France}
\affil[2]{Lebanese University, Computer Science Department, Beirut, Lebanon}
\affil[3]{ICB UMR 6303 CNRS - Univ. Bourgogne - Franche-Comté, Dijon, France}
\affil[4]{Univ Lyon, UCBL, CNRS, INSA Lyon, LIRIS, UMR5205, F-69622 Villeurbanne,  France}

\affil[*]{ali\_yassin@etu.u-bourgogne.fr}

\begin{abstract}

Networks are essential for analyzing complex systems. However, their growing size necessitates backbone extraction techniques aimed at reducing their size while retaining critical features. In practice, selecting, implementing, and evaluating the most suitable backbone extraction method may be challenging. This paper introduces {\fontfamily{qcr}\selectfont netbone}, a Python package designed for assessing the performance of backbone extraction techniques in weighted networks. Its comparison framework is the standout feature of {\fontfamily{qcr}\selectfont netbone}. Indeed, the tool incorporates state-of-the-art backbone extraction techniques. Furthermore, it provides a comprehensive suite of evaluation metrics allowing users to evaluate different backbones techniques. We illustrate the flexibility and effectiveness of {\fontfamily{qcr}\selectfont netbone} through the US air transportation network analysis. We compare the performance of different backbone extraction techniques using the evaluation metrics. We also show how users can integrate a new backbone extraction method into the comparison framework. {\fontfamily{qcr}\selectfont netbone} is publicly available as an open-source tool, ensuring its accessibility to researchers and practitioners. Promoting standardized evaluation practices contributes to the advancement of backbone extraction techniques and fosters reproducibility and comparability in research efforts. We anticipate that {\fontfamily{qcr}\selectfont netbone} will serve as a valuable resource for researchers and practitioners enabling them to make informed decisions when selecting backbone extraction techniques to gain insights into the structural and functional properties of complex systems.

\end{abstract}
\begin{document}

\flushbottom
\maketitle
%
%
\thispagestyle{empty}


\section*{Introduction}
    In recent years, the exponential growth of available data has prompted a surge in studying complex systems across various research domains. Networks have become a standard tool for modeling the entities and their interactions within such systems, with nodes and edges representing the entities and their relationships, respectively~\cite{complex:networks1, complex:networks10, complex:networks2, complex:networks3, complex:networks4, complex:networks5, complex:networks6, complex:networks7, complex:networks8, complex:networks9}. Moreover, the toolbox for network analysis continues to expand, with the introduction of numerous tools to facilitate various network analysis tasks~\cite{tool1, tool10, tool11, tool12, tool13, tool14, tool15, tool16, tool17, tool18, tool19, tool2, tool20, tool21, tool22, tool23, tool24, tool3, tool4, tool5, tool6, tool7, tool8, tool9}. However, analyzing large networks can be challenging. One solution to this issue is reducing the network size while retaining its essential properties. This objective is an active research area referred with various terms in the literature, such as sparsification, summarization, validated network extraction, skeleton extraction, and backbone extraction ~\cite{backbone1, backbone10, backbone11, backbone12, backbone13, backbone14, backbone15, backbone16, backbone17, backbone18, backbone19, backbone2, backbone20, backbone21, backbone22, backbone23, backbone24, backbone25, backbone3, backbone4, backbone5, backbone6, backbone7, backbone9, backbone27, backbone26, backbone31, backbone28, backbone29, backbone30, backbone32, backbone33, backbone34}.

Backbone extraction offers several advantages, including reduced data volume and storage, faster graph algorithms and queries, support for interactive analysis, and noise elimination. Backbone extraction has a wide range of applications. One uses it for various types of networks, such as social~\cite{social2, social3, social4, social5, social6, social7, social8, social9, social10}, biological~\cite{herb1}, brain~\cite{brain1, brain2, brain3, brain4}, gene~\cite{gene1, gene2, gene3, gene4, gene5}, metabolic~\cite{metabolic1, metabolic2, metabolic3, metabolic4, metabolic5},  
food web~\cite{food1, food2}, environmental~\cite{env1, env2}, finance~\cite{finance1, finance2, finance3, finance4}, trade~\cite{trade1, trade2, trade3, trade4, trade5}, information~\cite{information1, information2, information3, information4}, political~\cite{political1, political2}, transportation~\cite{transportation1, transportation2, transportation3, transportation4, transportation5}
, and others~\cite{other1, other2, other3, other4}. These applications have a broad range of uses, including clustering, classification, community detection, outlier detection, pattern set mining, identification of sources of infection in large graphs, and visualization, among others.

Practitioners must operate the most suitable method for their various applications or use cases. Therefore, there has been a growing interest in comparing backbone extraction techniques in the literature~\cite{survey1, survey2, survey3, survey4, survey5, survey6}. Furthermore, new tools have been introduced to fulfill the multitude of applications requirements~\cite{backbone1, tool4}. These tools implement a variety of backbone extraction techniques in different frameworks.

One can distinguish two main backbone extraction approaches: structural and statistical.
Structural methods focus on the network's topological features to extract a backbone with specific structural properties. They remove nodes or edges less critical for the properties to preserve. In contrast, statistical methods aim at eliminating noisy nodes or edges that blur the network information. They evaluate the significance of nodes and edges using a hypothesis-testing framework. They remove nodes or links qualified as noise.

The tool introduced by Coscia in~\cite{backbone1} incorporates three statistical and three structural backbone-extracting methods for weighted networks. Its {\fontfamily{qcr}\selectfont Python} module uses pandas~\cite{pandas} to enhance the performance. The backbone extraction techniques operate on a {\fontfamily{qcr}\selectfont DataFrame} input. They can process directed and undirected networks.

Neal presents {\fontfamily{qcr}\selectfont backbone} an {\fontfamily{qcr}\selectfont R} package to extract network backbones in~\cite{tool4}. It implements seventeen backbone extraction methods. Six methods are primarily designed for bipartite projections, two for weighted networks, and ten for unweighted networks. It also provides the generic sparsify() function that allows the custom construction of many more backbone methods. The methods operate on a {\fontfamily{qcr}\selectfont R Matrix} object, sparse {\fontfamily{qcr}\selectfont Matrix} object, a {\fontfamily{qcr}\selectfont DataFrame} object, or an {\fontfamily{qcr}\selectfont igraph} as input. They allow the processing of directed and undirected networks. 


Traditionally, users need to implement their code to compare different backbone extraction methods. We introduce {\fontfamily{qcr}\selectfont netbone}, a Python package specifically designed for extracting and comparing backbones from simple weighted networks. It offers an extensive collection of methods, including six statistical, thirteen structural, and one hybrid backbone extraction methods. It also provides filtering flexibility to tailor the backbone extraction process. Furthermore, it implements multiple ways to compare backbones.

They include a boolean filter for extraction techniques that extract a single backbone. Threshold and fraction filters are dedicated to methods assigning scores to the nodes or links. Users can indicate a threshold value to filter out elements with scores below its value. They can also show the fraction of features preserved associated with the top scores. In addition, the package provides a comparison framework with a visualization module to plot the results. Its goal is to assist the users in comparing various backbone extraction methods using a set of evaluation measures. This framework allows comparing backbone properties, distributions, and the evolution of various network properties when the backbone size is tunable. The package includes a set of predefined properties used for evaluating the extracted backbones. Moreover, users can integrate their backbone extraction methods and evaluation measures into the comparison framework. Furthermore, the framework facilitates the extraction of the consensual backbone, characterized by the common nodes and edges among a given set of backbones.

In the following sections, first, we briefly introduce the backbone extraction methods implemented in the {\fontfamily{qcr}\selectfont netbone} package. Then we present the package architecture and its modules, highlighting its numerous advantages. Next, we provide a simple toy example illustrating how to use {\fontfamily{qcr}\selectfont netbone}. Finally, we showcase the power of {\fontfamily{qcr}\selectfont netbone}'s comparison framework through five experiments. The first experiment illustrates how the comparison framework can assist in evaluating the backbone extraction methods by comparing various topological properties. The second experiment highlights how the framework could aid users in determining the appropriate fraction or threshold for extracting backbones. The third experiment illustrates how users evaluate the distribution of property values of the extracted backbones. The fourth experiment introduces the consensual backbone and how users can create unlimited combinations using the backbone extraction methods. Finally, the fifth experiment illustrates how users can integrate their new backbone extraction method into the comparison framework and evaluate it using their custom evaluation measures.

\section*{Backbone extraction methods}
\label{sec:methods}

Backbone extraction methods identify the most significant or essential parts within a network. Edge filtering techniques capture the most important relationships between nodes while removing less meaningful or noise-like connections. However, defining a crucial link in a network can be subjective and dependent on the specific application or research question. Therefore, researchers have developed several approaches to identify and extract the backbone of a network, each with its assumptions and criteria for importance. One can distinguish mainly statistical and structural methods. Additionally, hybrid methods incorporate both statistical and structural approaches. Table~\ref{methods:table} summarizes the main features of methods implemented in {\fontfamily{qcr}\selectfont netbone}.

    \subsection*{Statistical backbone extraction methods}
    Statistical backbone methods evaluate the significance of edges in a network using hypothesis testing based on empirical distribution or a null model. They compute p-values for each edge and filter edges based on their p-values. {\fontfamily{qcr}\selectfont netbone} implements six statistical backbone filtering techniques:
    \begin{itemize}

        \item \textbf{Disparity Filter~\cite{backbone2}:} It assumes that the normalized weights of a node's edges follow a uniform distribution. Comparisons of the observed normalized edge weights to this null model allow filtering out edges at a desired significance level $\alpha$. Since we define a null model for each node, an edge weight can be significant from the viewpoint of one of its nodes and not the other.

        \item \textbf{Marginal Likelihood Filter~\cite{backbone3}:} While the Disparity Filter assess the significance of an edge in the light of each node it connects independently, the Marginal Likelihood filter considers the two nodes the edge connects. It assumes an integer-weighted link as multiple unit edges. The null model assumes that each unit edge randomly chooses two nodes, which results in a binomial distribution. In other words, it calculates the probability of drawing at least $w$ unit edges from the strength of the network (summation of all weights) with probability proportional to both nodes' strengths. 

        \item \textbf{Noise Corrected Filter~\cite{backbone1}:} Like the Marginal Likelihood Filter, it assumes edge weights are drawn from a binomial distribution. However, using a Bayesian framework, it estimates the probability of observing a weight connecting two nodes. This framework enables us to generate posterior variances for all edges. This posterior variance allows us to create a confidence interval for each edge weight. Finally, we remove an edge if its weight is less than $\delta$ standard deviations stronger than the expectation ($\delta$ is the only parameter of the algorithm). It also provides a direct approximation through Binomial distribution similar to the Marginal Likelihood Filter.
        
        \item \textbf{Enhanced Configuration Model Filter~\cite{backbone25}:} It Enhances the null model of the Marginal Likelihood filter. Using the Enhanced Configuration Model of network reconstruction, its null model is based on the canonical maximum-entropy ensemble of weighted networks with the same degree and strength distribution as the actual network.

        \item \textbf{Locally Adaptive Network Sparsification Filter~\cite{backbone6}:} It makes no assumptions about the underlying weight distribution. Instead, the empirical cumulative density function is used to evaluate the statistical significance. Thus, from the viewpoint of edge incident nodes, it calculates the probability of choosing an edge randomly with a weight equal to the observed weight.

        \item \textbf{Multiple Linkage Analysis~\cite{backbone26}:} It assumes that the weights are evenly distributed among the node's neighbors, ranging from 1 to n. It calculates a goodness of fit by comparing the observed distribution with the hypothetical ones using a correlation coefficient. The optimal number of edges to retain for each node is determined by the number that yields the highest correlation coefficient.

    \end{itemize}
    
    \subsection*{Structural backbone extraction methods}
    Structural backbone methods operate on the network's topology to extract a backbone with specific topological properties. One can divide them into two categories. The first category includes techniques for extracting a single substructure from the network. They cannot be adjusted and typically result in a single backbone. The second category assigns scores to nodes or edges based on topological features. These methods can be tuned by setting a threshold $\beta$ or selecting the top fraction of scores. {\fontfamily{qcr}\selectfont Netbone} contains thirteen structural backbone filtering techniques:
       \begin{itemize}
       \item \textbf{Global Threshold Filter:} It is the most straightforward technique. It filters edges with weights lower than a predefined threshold $\beta$. 
       
        \item \textbf{Maximum Spanning Tree Filter:} It extracts a subgraph that includes all the nodes connected without forming cycles with the maximum total edge weight. 

        \item \textbf{Doubly Stochastic Filter~\cite{backbone5}:} It transforms the network's adjacency matrix into a doubly stochastic matrix by iteratively normalizing the row and column values using their respective sums. Next, one sorts the edges in descending order based on their normalized weight. One adds the edges to the backbone sequentially until it includes all nodes in the original network as a single connected component. It is not always possible to transform the matrix into a doubly-stochastic one.

        \item \textbf{High Salience Skeleton Filter~\cite{backbone7}:} It is based on the concept of edge salience. First, one constructs a shortest path tree for each node by merging all the shortest paths from that node to every other node in the network. Then the edge salience is computed as the proportion of shortest-path trees where the edge is present. The authors observed that edge salience follows a bimodal distribution near the boundaries 0 and 1. Consequently, they retain only the edges with salience near 1, eliminating the need to select an arbitrary threshold. 

        \item \textbf{h-Backbone Filter~\cite{backbone16}:} It is inspired by the h-index and edge betweenness. First, using the edge weights, it extracts the h-strength network: h is the largest natural number such that there are h links, each with a weight at least equal to h. Then it extracts the h-bridge network similarly. A bridge of an edge is the edge betweenness divided by the number of all nodes. Finally, the h-backbone merges the two networks.

        \item \textbf{Metric and Ultrametric distance backbone filters~\cite{backbone19}:} Both methods extract a subgraph consisting of the shortest paths in the network. Still, they diverge in their definitions of the shortest path length. Specifically, the Metric filter defines the shortest path length as the sum of the edge distances. In contrast, the Ultrametric filter defines it as the maximum distance among all edges in the path.

        \item \textbf{Modularity Backbone filter~\cite{backbone17}:} It is based on the concept of the Vitality Index. The Vitality Index measures the contribution of a node to the network's modularity. It computes the modularity variation before and after removing a network node. One extracts the backbone by setting a threshold value on the node's vitality index or selecting a top vitality fraction of nodes.

        \item \textbf{Planar Maximally Filtered Graph~\cite{backbone29}:} It simply reconstructs the graph by adding edges with the highest weight iteratively as long as the resulting graph is still planar.

        \item \textbf{Primary Linkage Analysis~\cite{backbone31}:} This method preserves the edge with the largest weight for each node.

        \item \textbf{Global Sparsification~\cite{backbone32}:} Assuming that an edge is likely to be within the same cluster if the nodes at its endpoints have a high neighbor overlap, the algorithm calculates the similarity of the endpoints using the Jaccard similarity. Then, it extracts the backbone of the graph, applying a threshold to the edge similarity.

        \item \textbf{Node Degree~\cite{backbone30}:} Node degree is computed by counting the number of connections or links a node has with other nodes in the network. One can filter the nodes based on their degree scores by setting a threshold.

        \item \textbf{Edge Betweenness~\cite{backbone28}:} It computes the edge betweenness of each edge. First, it finds the shortest paths in the network. Then, for each edge, it counts the shortest paths passing through it. Edges can be filtered using a threshold based on their edge betweenness scores.

    \end{itemize}

\subsection*{Hybrid backbone extraction methods}
The hybrid backbone extraction methods offer a unique approach by combining statistical and structural methodologies. These methods first calculate edge or node scores based on the network's topology. Subsequently, a statistical test is applied to these computed scores.
\begin{itemize}
    \item \textbf{Globally and Locally Adaptive Backbone~\cite{backbone33}:} It combines the Disparity and High Salience Skeleton filters. It measures the involvement of an edge by the fraction of all the shortest paths connecting a node to the rest of the network through this edge. The edge involvement is computed at the node level. Furthermore, one uses a null hypothesis to determine the statistical significance of each edge based on its involvement. The involvement is assumed to follow a uniform Gaussian or power law distribution. A parameter regulates the influence of the node's degree on its statistical significance.
\end{itemize}

\begin{table}[ht]
        \centering
            \def\arraystretch{1.2}
            \begin{tabular}{|c|c|c|c|c|c|c|}
                \hline\rule{0pt}{12pt} \multirow{2}{*}{\textbf{Category}} & 
                \multirow{2}{*}{\textbf{Method}} & \multicolumn{2}{c|}{\textbf{Network}} & \multicolumn{2}{c|}{\textbf{Filter}} & \multirow{2}{*}{\textbf{Parameters}}\\
                 \cline{3-6} & & \rotatebox{90}{Weighted} & \rotatebox{90}{Unweighted}& Type& Scope&\\
                \hline \multirow{6}{*}{Statistical} & Disparity & \textcolor{teal}{\cmark}& \textcolor{purple}{\xmark}& Edges& Local& alpha (significance level) \\
                 \cline{2-6} & Noise Corrected & \textcolor{teal}{\cmark}& \textcolor{purple}{\xmark}& Edges& Local& alpha (significance level)\\
                 \cline{2-6} & Marginal Likelihood & \textcolor{teal}{\cmark}& \textcolor{purple}{\xmark}& Edges& Local& alpha (significance level)\\
                 \cline{2-6} & Enhanced Configuration Model & \textcolor{teal}{\cmark}& \textcolor{purple}{\xmark}& Edges& Local& alpha (significance level)\\
                 \cline{2-6} & Locally Adaptive Network Sparsification & \textcolor{teal}{\cmark}& \textcolor{purple}{\xmark}& Edges& Local& alpha (significance level)\\
                 \cline{2-6} & Multiple Linkage Analysis & \textcolor{teal}{\cmark}& \textcolor{purple}{\xmark}& Edges& Local& - \\
                 
                 \hline \multirow{13}{*}{Structural} & Global threshold & \textcolor{teal}{\cmark}& \textcolor{purple}{\xmark} & Edges& Global& threshold\\
                 \cline{2-6} & Maximum Spanning Tree & \textcolor{teal}{\cmark}& \textcolor{purple}{\xmark} & Edges& Global& - \\
                 \cline{2-6} & Doubly Stochastic & \textcolor{teal}{\cmark}& \textcolor{purple}{\xmark} & Edges& Local& threshold\\
                 \cline{2-6} & High Salience Skeleton & \textcolor{teal}{\cmark}& \textcolor{purple}{\xmark} & Edges& Global& threshold\\
                 \cline{2-6} & h-Backbone & \textcolor{teal}{\cmark}& \textcolor{purple}{\xmark} & Edges& Global& -\\
                 \cline{2-6} & Metric Distance Backbone & \textcolor{teal}{\cmark}& \textcolor{purple}{\xmark} & Edges& Global& - \\
                 \cline{2-6} & Ultrametric Distance Backbone & \textcolor{teal}{\cmark}& \textcolor{purple}{\xmark}& Edges& Global& - \\
                 \cline{2-6} & Planar Maximally Filtered Graph & \textcolor{teal}{\cmark}& \textcolor{purple}{\xmark}& Edges& Global& - \\
                 \cline{2-6} & Modularity Backbone & \textcolor{teal}{\cmark}& \textcolor{purple}{\xmark}& Nodes& Global& threshold \\
                 \cline{2-6} & Primary Linkage Analysis & \textcolor{teal}{\cmark}& \textcolor{purple}{\xmark}& Edges& Local& - \\
                 \cline{2-6} & Global Sparsification & \textcolor{teal}{\cmark}& \textcolor{teal}{\cmark}& Edges& Local& threshold \\
                \cline{2-6} & Edge Betweenness & \textcolor{teal}{\cmark}& \textcolor{teal}{\cmark}& Edges& Global& threshold \\
                \cline{2-6} & Node Degree & \textcolor{teal}{\cmark}& \textcolor{teal}{\cmark}& Nodes& Global& threshold \\
                 
                 \hline Hybrid & Globally and Locally Adaptive Backbone & \textcolor{teal}{\cmark}& \textcolor{teal}{\cmark}& Edges& Local \& Global& \shortstack[c]{c (involvement parameter) \\ alpha (significance level) }\\
                \hline
            \end{tabular}
             \caption{\textbf{A summary of the backbone extraction method characteristics implemented in netbone. Including network types (weighted/unweighted), filter type and scope (edges/nodes and local/global), and method parameters. \textcolor{teal}{\cmark} indicate the applicability and \textcolor{purple}{\xmark} indicate the inapplicability}}
             \label{methods:table}
    \end{table}

\section*{The NetBone Package}
\label{sec:netbone}

\begin{figure}[ht]
\centering
\includegraphics[width=\linewidth]{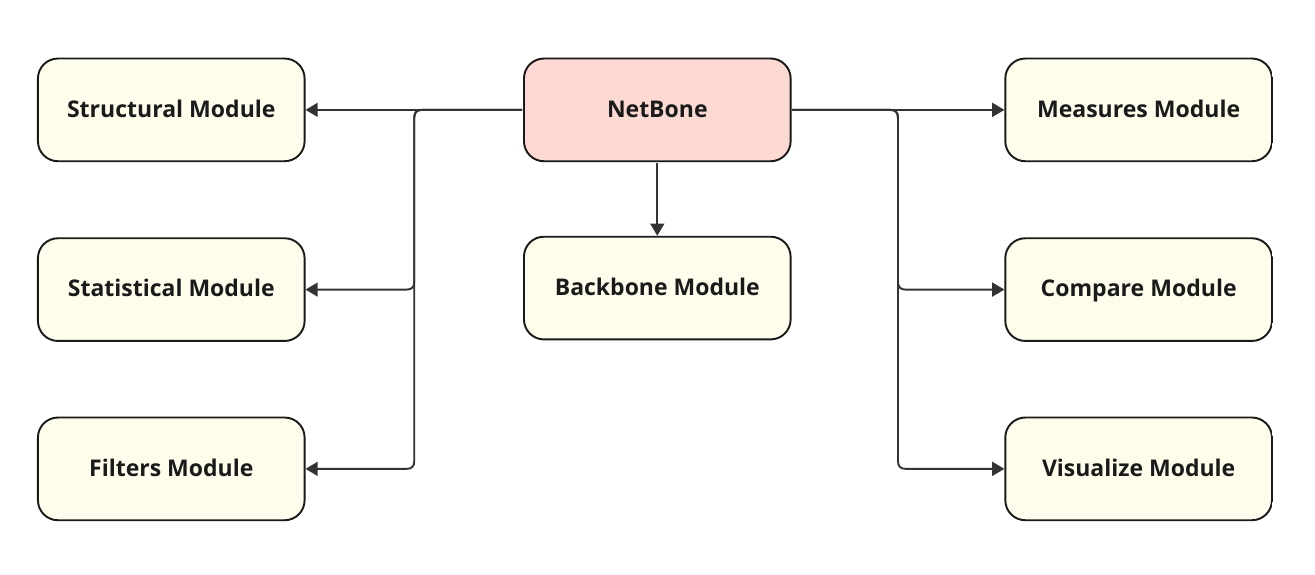}
\caption{The diagram illustrates the different modules provided in the {\fontfamily{qcr}\selectfont netbone} package and their interactions.}
\label{fig:diagram}
\end{figure}

{\fontfamily{qcr}\selectfont Netbone} is a Python package freely available to the public on GitLab (\url{https://gitlab.liris.cnrs.fr/coregraphie/netbone}). It provides a straightforward and easy-to-use framework for comparing and selecting the most appropriate method for a given case study. Figure~\ref{fig:diagram} provides a diagram representing the various modules of the {\fontfamily{qcr}\selectfont netbone} package. We give a brief presentation of its architecture and present its various modules with their main features.

\begin{itemize}
    \item \textbf{Backbone Module:} The {\fontfamily{qcr}\selectfont Backbone} module contains the {\fontfamily{qcr}\selectfont Backbone} class, which is central to the backbone extraction functionality. Running a backbone extraction method returns an instance of this class. It contains the calculated scores (for structural methods) or p-values (for statistical methods) associated with the nodes or edges of the chosen backbone extraction technique. One can inspect the scores or p-values by invoking the {\fontfamily{qcr}\selectfont to\_dataframe()} function. It generates a data frame with the corresponding scores or p-values. Additionally, the module allows users to easily incorporate their newly defined backbone extraction method into the {\fontfamily{qcr}\selectfont netbone} comparison framework. All that is required is for the user to return an instance of the {\fontfamily{qcr}\selectfont Backbone} class at the end of their function.
    
    \item \textbf{Structural and Statistical Modules:} The {\fontfamily{qcr}\selectfont statistical} and {\fontfamily{qcr}\selectfont structural} modules group methods based on their underlying methodology. Invoking a function from these modules calculates p-values for statistical methods such as the {\fontfamily{qcr}\selectfont disparity\_filter()}. It computes scores for some structural techniques such as the {\fontfamily{qcr}\selectfont high\_salience\_skeleton()}. For structural methods that extract a substructure from the graph, such as the {\fontfamily{qcr}\selectfont maximum\_spanning\_tree()}, it assigns Boolean values to the edges or nodes of the network. In all cases, it returns a new instance of the {\fontfamily{qcr}\selectfont Backbone} class containing the computed values.
    
    \item \textbf{Filters Module:} The {\fontfamily{qcr}\selectfont filters} module is a powerful component allowing users to extract the backbone that meets their specific needs. It accomplishes this by separating the backbone values calculation (score, p-value, Boolean) and the filtering process. For example, users can use the {\fontfamily{qcr}\selectfont threshold\_filter()} function to extract a backbone based on a score or p-value threshold. They can use the {\fontfamily{qcr}\selectfont fraction\_filter()} function to obtain a backbone of a desired size. The {\fontfamily{qcr}\selectfont boolean\_filter()} function is designed for methods extracting a single substructure of the network. Table~\ref{filters:table} summarizes the filters associated with the various backbone extraction methods.
    
    \item \textbf{Measures Module:}The {\fontfamily{qcr}\selectfont measures} module contains a set of evaluation measures allowing users to compute the topological properties of the extracted backbone. These measures have been carefully defined. They include those used in the seminal work of Serrano~\cite{backbone2}.

    \item \textbf{Compare Module:} The {\fontfamily{qcr}\selectfont compare} module is a standout feature, providing a robust comparison framework through its  {\fontfamily{qcr}\selectfont Compare} class. The module offers four main functions. First, the {\fontfamily{qcr}\selectfont properties()} function computes a set of specified properties of the original network and the extracted backbones using the desired Filter. Second, the {\fontfamily{qcr}\selectfont properties\_progression()} function computes the evolution of given properties between the original network and a set of extracted backbones. This set can be defined using a filter type and an array of thresholds or fractions. Third, the {\fontfamily{qcr}\selectfont distribution\_ks\_statistic()} function computes the KS statistic~\cite{ks:statistic} between the cumulative distribution of an original network property and its distribution in a given backbone. It needs a function to extract the property values and a filter type. Finally, using the {\fontfamily{qcr}\selectfont consent()} functions, the module allows the extraction of what is called the consensual backbone. Given a set of different backbones, the method computes the intersection between the backbones
    
    \item \textbf{Visualize Module:} The {\fontfamily{qcr}\selectfont visualize} module is designed to facilitate the comparisons by generating visually appealing plots. It contains three main plotting functions. First, the {\fontfamily{qcr}\selectfont radar\_plot()} function generates a radar chart with a separate axis for each added property. This chart is useful as it clearly and concisely represents multiple properties on a single plot. It simplifies the backbone's analysis and comparison across various topological dimensions. Second, the {\fontfamily{qcr}\selectfont progression\_plot()}function generates simple line charts that display the evolution of defined properties as the fraction or threshold changes. Finally, the {\fontfamily{qcr}\selectfont distribution\_plot()} function generates simple scatter charts that display the distribution of defined property values in the extracted backbones.

\end{itemize}
\begin{table}[ht]
        \begin{center}
            \def\arraystretch{1.2}
            \begin{tabular}{|c|c|c|c|}
                \hline
                \textbf{Backbone Extraction Method} & \textbf{Boolean Filter} & \textbf{Fraction Filter} & \textbf{Threshold Filter}\\
                \hline
                Disparity & \textcolor{purple}{\xmark} & \textcolor{teal}{\cmark} & \textcolor{teal}{\cmark}\\
                \hline
                Noise Corrected & \textcolor{purple}{\xmark} & \textcolor{teal}{\cmark} & \textcolor{teal}{\cmark}\\
                \hline
                Marginal Likelihood & \textcolor{purple}{\xmark} & \textcolor{teal}{\cmark} & \textcolor{teal}{\cmark}\\
                \hline
                Enhanced Configuration Model & \textcolor{purple}{\xmark} & \textcolor{teal}{\cmark} & \textcolor{teal}{\cmark}\\
                \hline
                Locally Adaptive Network Sparsification & \textcolor{purple}{\xmark} & \textcolor{teal}{\cmark} & \textcolor{teal}{\cmark}\\
                \hline
                Multiple Linkage Analysis & \textcolor{teal}{\cmark} & \textcolor{purple}{\xmark} & \textcolor{purple}{\xmark}\\
                \hline
                Global threshold & \textcolor{purple}{\xmark} & \textcolor{teal}{\cmark} & \textcolor{teal}{\cmark}\\
                \hline
                Maximum Spanning Tree & \textcolor{teal}{\cmark} & \textcolor{purple}{\xmark} & \textcolor{purple}{\xmark}\\
                \hline
                Doubly Stochastic & \textcolor{teal}{\cmark} & \textcolor{teal}{\cmark} & \textcolor{teal}{\cmark}\\
                \hline
                High Salience Skeleton & \textcolor{teal}{\cmark} & \textcolor{teal}{\cmark} & \textcolor{teal}{\cmark}\\
                \hline
                h-Backbone & \textcolor{teal}{\cmark} & \textcolor{purple}{\xmark} & \textcolor{purple}{\xmark}\\
                \hline
                Metric Distance Backbone & \textcolor{teal}{\cmark} & \textcolor{purple}{\xmark} & \textcolor{purple}{\xmark}\\
                \hline
                Ultrametric Distance Backbone & \textcolor{teal}{\cmark} & \textcolor{purple}{\xmark} & \textcolor{purple}{\xmark}\\
                \hline
                Modularity Backbone  & \textcolor{purple}{\xmark} & \textcolor{teal}{\cmark} & \textcolor{teal}{\cmark}\\
                \hline
                Planar Maximally Filtered Graph & \textcolor{teal}{\cmark} & \textcolor{purple}{\xmark} & \textcolor{purple}{\xmark}\\
                \hline
                Primary Linkage Analysis & \textcolor{teal}{\cmark} & \textcolor{purple}{\xmark} & \textcolor{purple}{\xmark}\\
                \hline
                Global Sparsification  & \textcolor{purple}{\xmark} & \textcolor{teal}{\cmark} & \textcolor{teal}{\cmark}\\
                \hline
                Edge Betweenness & \textcolor{purple}{\xmark} & \textcolor{teal}{\cmark} & \textcolor{teal}{\cmark}\\
                \hline
                Node Degree & \textcolor{purple}{\xmark} & \textcolor{teal}{\cmark} & \textcolor{teal}{\cmark}\\
                \hline
                Globally and Locally Adaptive Backbone & \textcolor{purple}{\xmark} & \textcolor{teal}{\cmark} & \textcolor{teal}{\cmark}\\
                \hline
            \end{tabular}
             \caption{\textbf{A summary of the filters that can(\textcolor{teal}{\cmark}) and cannot(\textcolor{purple}{\xmark}) be applied for each backbone extraction method.}}
             \label{filters:table}
        \end{center}
    \end{table}

\section*{NetBone in action: A toy example}
To illustrate the usage of {\fontfamily{qcr}\selectfont netbone}, we consider the high salience skeleton method with the Les Mis{\'e}rables network~\cite{lesmis}. We chose this extraction technique because it can be associated with the three filtering methods provided by {\fontfamily{qcr}\selectfont netbone}. To begin using the {\fontfamily{qcr}\selectfont netbone} package, one can install the latest release from either the PyPI repository (\url{https://pypi.org/project/netbone/}) or directly from the project's GitLab repository:\\ \\
{\fontfamily{qcr}\selectfont > pip install netbone\\
> pip install git+https://gitlab.liris.cnrs.fr/coregraphie/netbone}\\\\
Once installed, the {\fontfamily{qcr}\selectfont netbone} package can be imported using:\\\\
{\fontfamily{qcr}\selectfont 
> import netbone as nb}\\\\
The {\fontfamily{qcr}\selectfont netbone} package can handle two types of inputs: a {\fontfamily{qcr}\selectfont networkx} graph or a {\fontfamily{qcr}\selectfont DataFrame}. In this example, we will load the Les Mis{\'e}rables network from {\fontfamily{qcr}\selectfont networkx} and apply the {\fontfamily{qcr}\selectfont high\_salience\_skeleton()} method. The resulting scores can be examined using the {\fontfamily{qcr}\selectfont to\_dataframe()} function as shown below:\\\\
{\fontfamily{qcr}\selectfont 
> import networkx as nx\\
> g = nx.les\_miserables\_graph()\\
> b = nb.high\_salience\_skeleton(g)\\
> b.to\_data\_frame()\\\\\\\\
| \ source \ | \ \ \ \ target \ \ \ \ | \ weight | high\_salience\_skeleton | \ score |\\
|-------------------|--------------------------------|------------------|------------------------------------------------|----------------|\\
| Napoleon | \ \ \ \ Myriel \ \ \ \ | \ \ \ 1 \ \ \ | \ \ \ \ \ \ \ \ \ True \ \ \ \ \ \ \ \ \ | \ 1.000  |\\
| \ Myriel \ | MlleBaptistine | \ \ \ 8 \ \ \ | \ \ \ \ \ \ \ \ \ True \ \ \ \ \ \ \ \ \ | \ 0.987  |\\
| \ Myriel \ | \  MmeMagloire \ \ | \ \ 10 \ \ \ | \ \ \ \ \ \ \ \ \ True \ \ \ \ \ \ \ \ \ | \ 0.987  |\\
| \ Myriel \ | \ CountessDeLo \ | \ \ \ 1 \ \ \ | \ \ \ \ \ \ \ \ \ True \ \ \ \ \ \ \ \ \ | \ 1.000  |\\
| \ Myriel \ | \ \ \ Geborand \ \ \ | \ \ \ 1 \ \ \ | \ \ \ \ \ \ \ \ \ True \ \ \ \ \ \ \ \ \ | \ 1.000  |\\
| \ Myriel \ | \ Champtercier \ | \ \ \ 1 \ \ \ | \ \ \ \ \ \ \ \ \ True \ \ \ \ \ \ \ \ \ | \ 1.000  |\\
| \ \ ... \ \ | \ \ \ \ \ ... \ \ \ \ \ | \ \ ... \ | \ \ \ \ \ \ \ \ \ \ ... \ \ \ \ \ \ \ \ | \ ... \ |\\
|-------------------|--------------------------------|------------------|------------------------------------------------|----------------|\\
}\\\\
The high salience skeleton method proposed by Grady exhibits a bimodal distribution of scores centered around 0 and 1. The default approach of this method is to keep only edges with scores greater than 0.8. In {\fontfamily{qcr}\selectfont netbone}, it can be accomplished using the {\fontfamily{qcr}\selectfont boolean\_filter()}. However, in that case, two nodes are missing from the extracted backbone in this particular example. To fix this issue, users can adjust the threshold by using the {\fontfamily{qcr}\selectfont threshold\_filter()} function. One can use a threshold of 0.7 to retain all the network nodes. Additionally, users can control the size of the backbone using the {\fontfamily{qcr}\selectfont fraction\_filter()}, such as keeping 15\% of the network. The following code shows how to do it in {\fontfamily{qcr}\selectfont netbone}:\\\\
{\fontfamily{qcr}\selectfont 
> from netbone.filters import  boolean\_filter, threshold\_filter, fraction\_filter\\
> backbone1 = boolean\_filter(b)\\
> backbone2 = threshold\_filter(b, 0.7)\\
> backbone3 = fraction\_filter(b, 0.15)}\\\\
Once backbones are extracted, users can use them in their applications and case studies. For the sake of simplicity, we visualize the backbones using {\fontfamily{qcr}\selectfont networkx} in a spiral layout. 

\begin{figure}[ht]
\centering
\includegraphics[width=\linewidth]{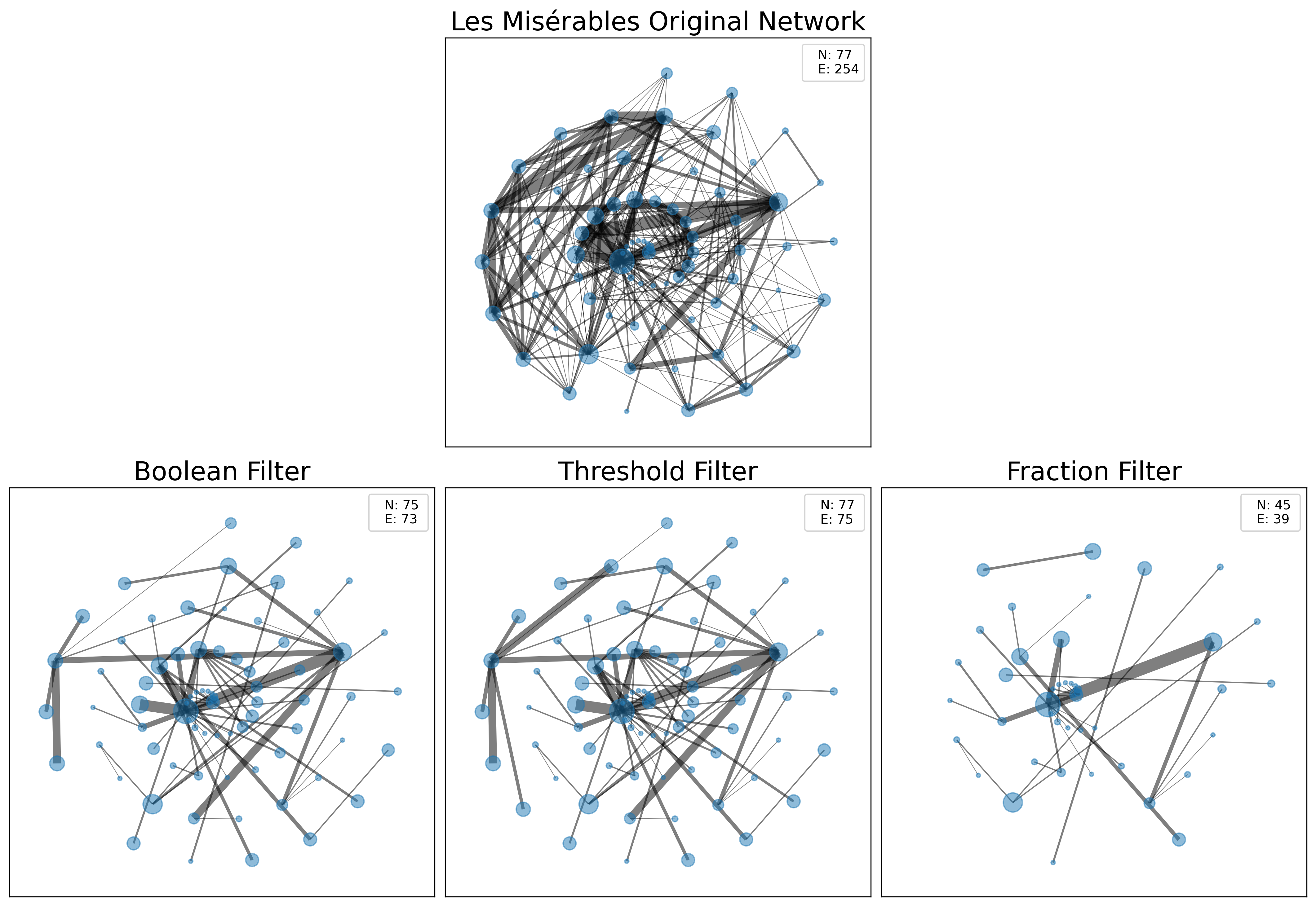}
\caption{The original Les Mis{\'e}rables network and its extracted backbones using the {\fontfamily{qcr}\selectfont boolean\_filter()} with a default threshold of 0.8, {\fontfamily{qcr}\selectfont threshold\_filter()} with a threshold value of 0.7, and {\fontfamily{qcr}\selectfont fraction\_filter()} with a fraction of 0.15. N and E are the number of nodes and edges, respectively. The size of the nodes is proportional to the degree. The width of the links is proportional to the weights.} 
\label{toy:example}
\end{figure}

Figure~\ref{toy:example} presents the Les Mis{\'e}rables original network and its backbones using the {\fontfamily{qcr}\selectfont boolean\_filter()}, {\fontfamily{qcr}\selectfont threshold\_filter()}, and {\fontfamily{qcr}\selectfont fraction\_filter()}. The size of the nodes is proportional to their degree, and the width of the link is proportional to their weights.
The lower-left panel of the figure displays the backbone processed with the {\fontfamily{qcr}\selectfont boolean\_filter()}. It is the output of the high-salience skeleton method with default values. It retains the edges that participate in at least 80\% of the shortest paths in the network. The backbone is sparser than the original network, with the fraction of links reduced by 70\%. However, some nodes are missing in this backbone. The lower-middle panel shows the backbone using the {\fontfamily{qcr}\selectfont threshold\_filter()} to adjust the threshold. It includes edges participating in at least 70\% of the shortest paths in the network. This backbone contains all the nodes and two more links than the previous one. The lower-right panel shows the backbone and {\fontfamily{qcr}\selectfont fraction\_filter()} retaining only the top 15\% scores of edges. It is the sparser backbone with multiple components. These edges connect 45 nodes and account for approximately 60\% of all nodes in the network.

\section*{Exploring NetBone's comparison framework}
The comparison framework of {\fontfamily{qcr}\selectfont netbone} stands out as a key feature. It allows users to easily explore and compare the backbones extracted from various methods with built-in evaluation measures. Moreover, users can easily integrate their backbone extraction methods and evaluation measures into the comparison framework. The framework provides five distinct use cases for comparison purposes:

The first use case involves comparing the backbone's topological properties. It is achieved using the {\fontfamily{qcr}\selectfont properties()} method to compute the selected properties of the backbones. One can visualize the results in a radar plot using the {\fontfamily{qcr}\selectfont radar\_plot()} method.
 
The second use case focuses on comparing the evolution of backbone properties. Users can compute the selected properties for various fractions of edges/nodes or varying significance levels or thresholds for the backbones. It can be done using the {\fontfamily{qcr}\selectfont properties\_progression()} method. One can visualize the results in line charts using the {\fontfamily{qcr}\selectfont progression\_plot()} method.

The third use case centers around comparing topological properties distributions. Users can assess the distances between the cumulative distributions for a given property and a couple of backbones. To do so, one must compute the Kolmogorov-Smirnov (KS) statistic of the cumulative distributions under evaluation using the {\fontfamily{qcr}\selectfont distribution\_ks\_statistic()} method. The {\fontfamily{qcr}\selectfont distribution\_plot()} method allows visualizing the differences in a scatter plot.

The fourth use case involves extracting the consensual backbone. It entails keeping identical nodes and edges among the backbones. It can be accomplished using the {\fontfamily{qcr}\selectfont consent()} method.

Finally, the fifth use case illustrates how users can integrate their backbone extraction methods and their custom evaluation measures into {\fontfamily{qcr}\selectfont netbone}'s comparison framework.

In the following subsections, we illustrate the ability of this framework to evaluate the effectiveness of backbone extraction methods across various applications. We use the US air transportation network introduced in the work of Serrano~\cite{backbone2}. It consists of 382 airport nodes in the continental US. The edges represent routes between these airports, and the weights assigned to the edges correspond to the number of passengers. The supplementary materials contain detailed explanations of the code for each experiment.

\subsection*{Experiment 1}
\begin{figure}[ht]
\centering
\includegraphics[width=\linewidth]{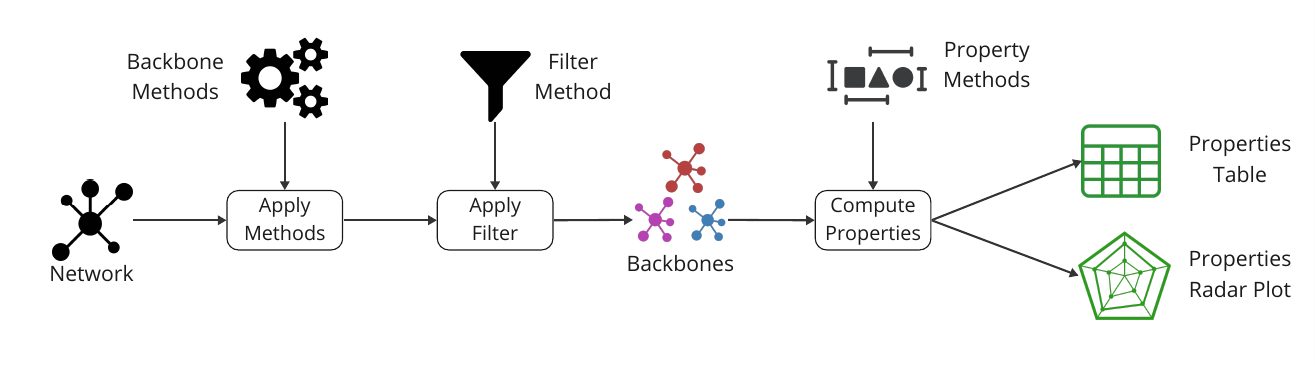}
\caption{A diagram illustrating the flow within {\fontfamily{qcr}\selectfont netbone}'s comparison framework to compute the topological properties of the extracted backbone. } 
\label{exp1}
\end{figure}

In this experiment, we focus on assessing the connectivity of the structural backbone extraction methods in the air transportation network using {\fontfamily{qcr}\selectfont netbone}'s comparison framework. The aim is to have a connected filtered network when applying filters since connectivity is an essential property in transportation networks. Figure~\ref{exp1} illustrates the process flow within {\fontfamily{qcr}\selectfont netbone}'s comparison framework for computing topological properties. 

First, we extract from the network backbones using eight structural backbone extraction methods. We use the Boolean Filter within the framework since these methods extract a substructure from the network. Next, we compute various properties, with a particular focus on reachability. Reachability measures the connectivity between nodes in a network by quantifying the fraction of node pairs that can communicate with each other. Furthermore, we examine additional properties such as node, edge, weight fractions, density, and average degree of the extracted backbones. The results are presented in a table for easy numerical analysis and can be visualized using a spider plot.

\begin{figure}[ht]
\centering
\includegraphics[width=\linewidth]{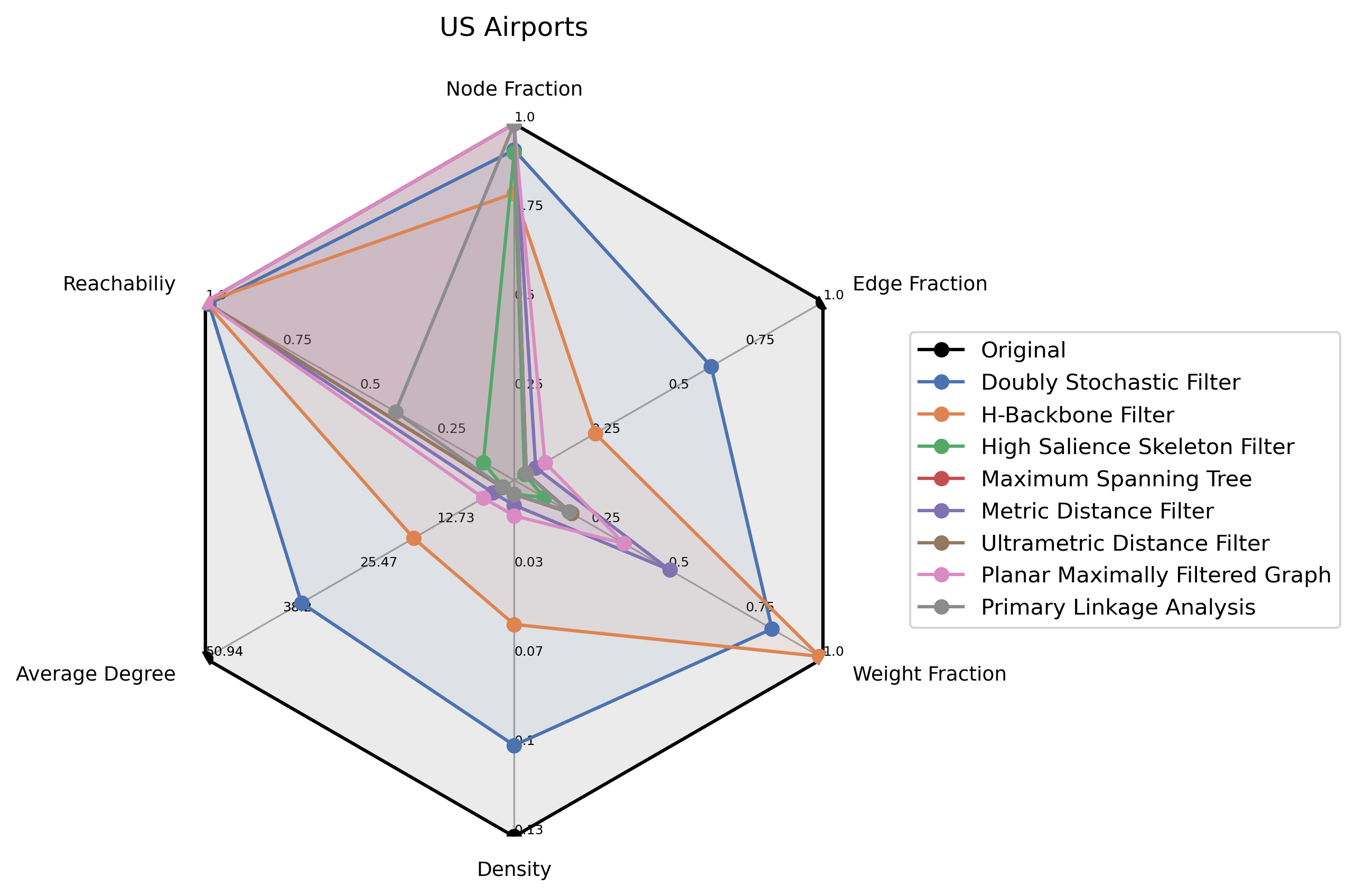}
\caption{A radar chart showing the topological properties of the extracted structural backbones plotted using {\fontfamily{qcr}\selectfont netbone}. The topological properties are the fraction of nodes, edges, and weights preserved in the backbone, density, average degree, and reachability of the extracted backbone.} 
\label{exp1.2}
\end{figure}

In Table~\ref{table:exp1} and Figure~\ref{exp1.2}, we present the topological properties of the extracted backbones. One can observe that all methods yield a backbone with a reachability value of 1, except for the Doubly Stochastic, Primary Link Analysis, and High Salience Skeleton methods. Reachability represents the fraction of node pairs that can communicate with each other in the network. Suppose that the user is interested in backbones with a reachability of 1. According to this criterion, one can exclude the Doubly Stochastic, Primary Link Analysis, and High Salience Skeleton methods.
 \begin{table}[ht]
        \begin{center}
            \def\arraystretch{1.2}
            \begin{tabular}{|c|c|c|c|c|c|c|}
                \hline\rule{0pt}{12pt}
                \textbf{Method} & \textbf{Reachability} & \textbf{Node Fraction} & \textbf{Edge Fraction} & \textbf{Weight Fraction} & \textbf{Density} & \textbf{Average Degree}\\
                \hline
                Original Network & 1 & 1 & 1 & 1 & 0.1344 & 50.9 \\
                \hline
                h-Backbone & 1 & 0.80 & 0.26 & 0.98 & 0.0544 & 16.5\\
                \hline
                Maximum Spanning Tree & 1 & 1 & 0.03 & 0.18 & 0.0053 & 1.99 \\
                \hline
                Metric Backbone & 1 & 1 & 0.06 & 0.50 & 0.0094 & 3.55 \\
                \hline
                Ultrametric Backbone & 1 & 1 & 0.03 & 0.18 & 0.0053 & 1.99 \\
                \hline
                Planar Maximally Graph & 1 & 1 & 0.09 & 0.35 & 0.0134 & 5.0 \\
                \hline
                Doubly Stochastic & 0.98 & 0.92 & 0.63 & 0.83 & 0.1 & 35.0 \\
                \hline
                Primary Linkage Analysis & 0.38 & 1 & 0.03 & 0.17 & 0.0052 & 1.9 \\
                \hline
                High Salience Skeleton & 0.1 & 0.91 & 0.03 & 0.09 & 0.0053 & 1.8 \\
                \hline
            \end{tabular}
            \caption{\textbf{The topological properties of the structural backbones computed using {\fontfamily{qcr}\selectfont netbone}'s comparison framework.}}
        \label{table:exp1}
        \end{center}
        \end{table}
Examining the Node fraction, we find that only the h-Backbone method isolates some nodes, as it preserves only 80\% of the nodes. Consequently, we exclude the h-Backbone from our selection.

Moving forward, among the remaining methods, our focus shifts to choosing the technique that preserves the highest weight fraction. Consequently, we select the Metric Backbone method, which retains 50\% of the weights. However, one can note that this method only includes 3\% of the edges, resulting in an average degree of 3.5 and a low density of 0.0094.
 
To summarize, this use case filters the original transportation network under constraints. Indeed, we aim to retain all nodes while ensuring they remain connected within a single component. Furthermore, one wants to maximize the preservation of weights. Using {\fontfamily{qcr}\selectfont netbone}'s comparison framework, we can evaluate and compare the performance of various backbone extraction methods based on these multiple properties. It allows us to identify the Metric Backbone method as it preserved the highest fraction of weights while maintaining connectivity within a single component.

\subsection*{Experiment 2}
\begin{figure}[ht]
\centering
\includegraphics[width=\linewidth]{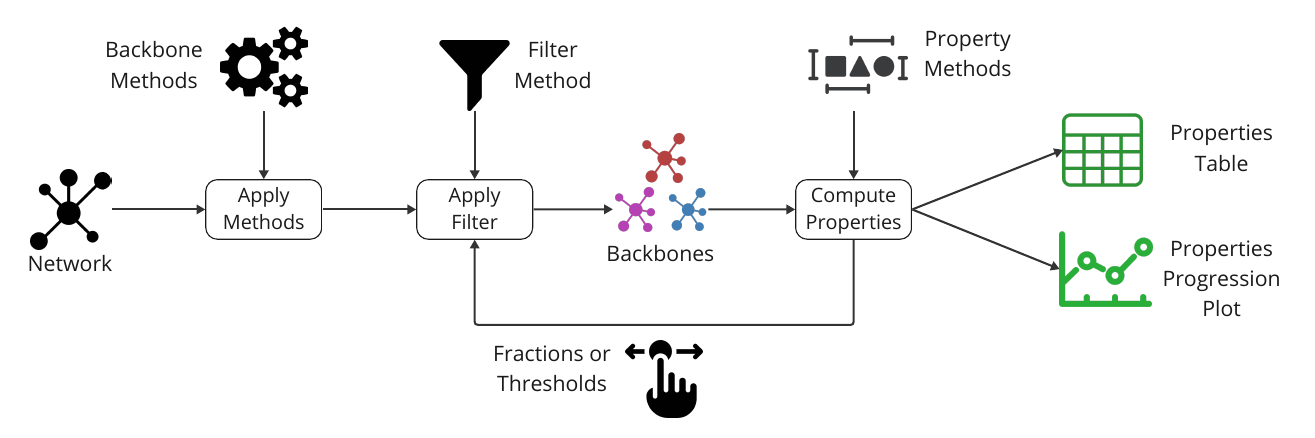}
\caption{A diagram illustrating the flow within {\fontfamily{qcr}\selectfont netbone} comparison framework to compute the evolution of the topological properties as the threshold varies in the extracted backbone.} 
\label{exp2}
\end{figure}

The Previous experiment focuses on the structural methods for backbone extraction. Some of these methods can be adjusted using a threshold on scores or selecting the top fraction of scores. In this experiment, our objective is to sparsify the network while preserving all the nodes, which is crucial in the context of a transportation network. To achieve this, we use {\fontfamily{qcr}\selectfont netbone}'s comparison framework to help us determine the appropriate fraction. Figure~\ref{exp2} illustrates the process flow within {\fontfamily{qcr}\selectfont netbone}'s comparison framework for computing topological properties as the fraction of edges or thresholds varies. 

In the experiment, we extract the backbones using five structural backbone extraction. Using the fraction filter, we gradually sparsify the network by adjusting the fraction from 0.01 to 0.5. We aim to keep the backbone edge size below 50\% of the original network. For each fraction, we compute the node fraction to assess the preservation of nodes. The results are in a table for easy analysis. Additionally, one can use a progression line plot to visualize the evolution of the node fraction as the fraction filter varies.

\begin{figure}[ht]
\centering
\includegraphics[width=\linewidth]{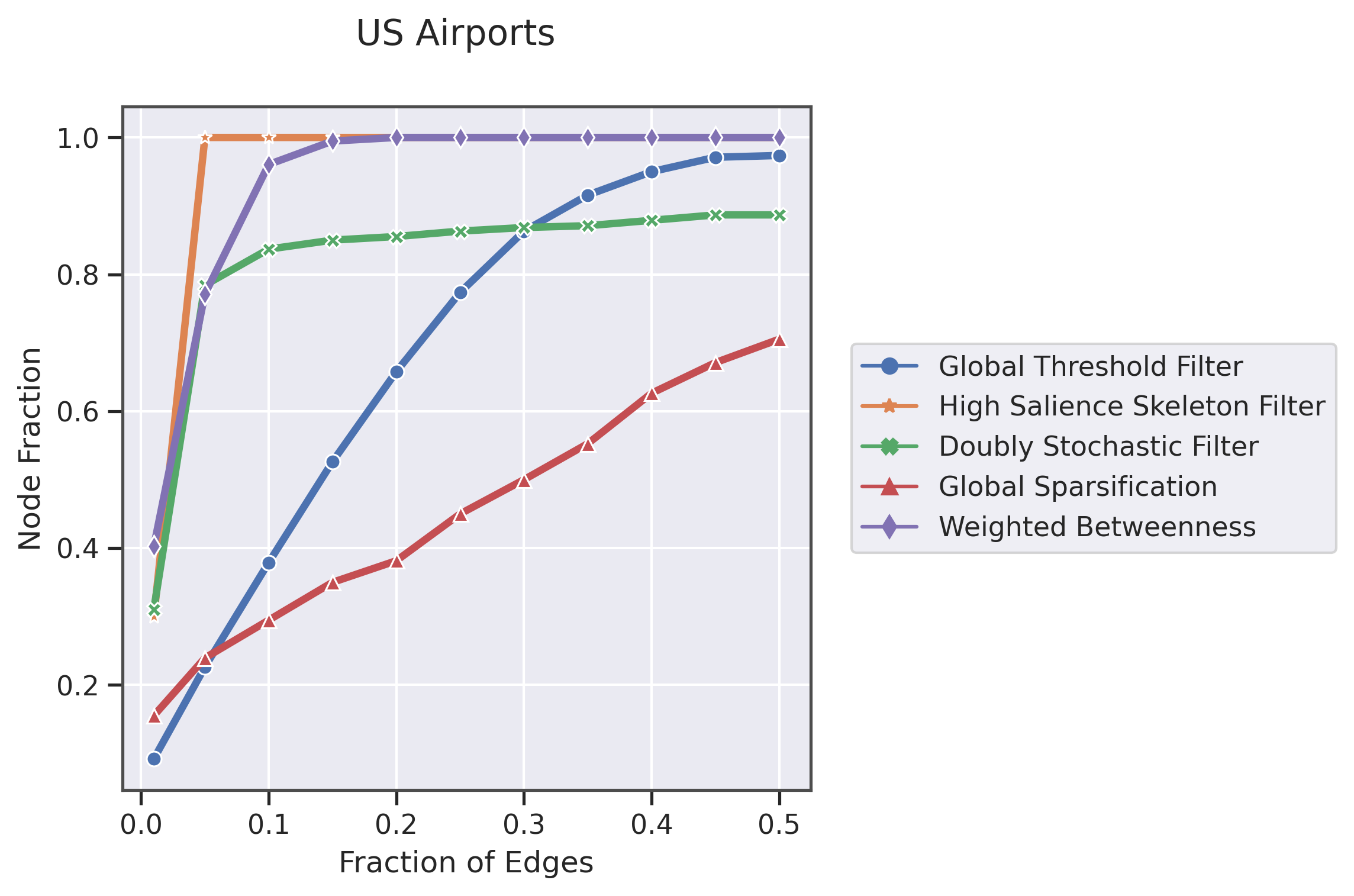}
\caption{A line chart showing the fraction of nodes of the extracted structural backbones as a function of the fraction of edges plotted using {\fontfamily{qcr}\selectfont netbone}.} 
\label{exp2.2}
\end{figure}

Table~\ref{table:exp2} and Figure~\ref{exp2.2} present the node fraction for each backbone extraction method as a function of the edge fraction. We observe that the global threshold, doubly stochastic, and global sparsification methods fail to extract a backbone that includes all the nodes while keeping the edge fraction below 50\%. Consequently, we can exclude these methods from our list of interests. However, the betweenness method allows us to maintain all the nodes with an edge fraction of 20\%. The high salience skeleton method stands out by enabling us to preserve all the nodes with an edge fraction as low as 5\%. 

\begin{table}[ht]
    \begin{center}
        \def\arraystretch{1.2}
        \begin{tabular}{|c|c|c|c|c|c|}
            \hline\rule{0pt}{12pt}
            \textbf{Edge Fraction} & \textbf{Global Threshold} & \textbf{High Salience} & \textbf{Doubly Stochastic} & \textbf{Global Sparsification} & \textbf{Weighted Betweenness} \\
            \hline
            0.01 & 0.09 & 0.3 & 0.31 & 0.15 & 0.40\\
            \hline
            0.05 & 0.22 & \cellcolor{lightgray} 1.0 & 0.78 & 0.23 & 0.77\\
            \hline
            0.10 & 0.37 & 1.0 & 0.83 & 0.29 & 0.96\\
            \hline
            0.15 & 0.52 & 1.0 & 0.85 & 0.35 & 0.99\\
            \hline
            0.20 & 0.65 & 1.0 & 0.85 & 0.38 & \cellcolor{lightgray}1.00\\
            \hline
            0.25 & 0.77 & 1.0 & 0.86 & 0.45 & 1.00\\
            \hline
            0.30 & 0.86 & 1.0 & 0.86 & 0.50 & 1.00\\
            \hline
            0.35 & 0.91 & 1.0 & 0.87 & 0.55 & 1.00\\
            \hline
            0.40 & 0.95 & 1.0 & 0.87 & 0.62 & 1.00\\
            \hline
            0.45 & 0.97 & 1.0 & 0.88 & 0.67 & 1.00\\
            \hline
        \end{tabular}
        \caption{\textbf{The node fraction for each backbone extraction method as a function of the edge fraction calculated using {\fontfamily{qcr}\selectfont netbone}'s comparison framework.}}
    \label{table:exp2}
    \end{center}
    \end{table}
This experiment aims to identify the optimal structural backbone extraction method to sparsify the transportation network while preserving all the nodes. Through {\fontfamily{qcr}\selectfont netbone}'s comparison framework, we evaluate different methods using the fraction filter. The high salience skeleton method successfully achieved the objective by retaining all nodes with a low edge fraction of 5\%. It is worth noting that one can use this approach with {\fontfamily{qcr}\selectfont netbone}'s comparison framework to other applications with alternative criteria.

\subsection*{Experiment 3}
In this experiment, we use {\fontfamily{qcr}\selectfont netbone}'s comparison framework to assess the global threshold and statistical methods to capture the weight and degree distributions. Indeed, using the global threshold method, the weight distribution is truncated. It emphasizes the edges between the hubs in the air transportation network. These edges typically have high weights due to the significant volume of passengers involving large carriers. One can use statistical methods to capture different scales of importance and highlight the hub and spoke topology. These methods account for multiple scales and provide a more comprehensive understanding of the network's structure.

\begin{figure}[ht]
\centering
\includegraphics[width=\linewidth]{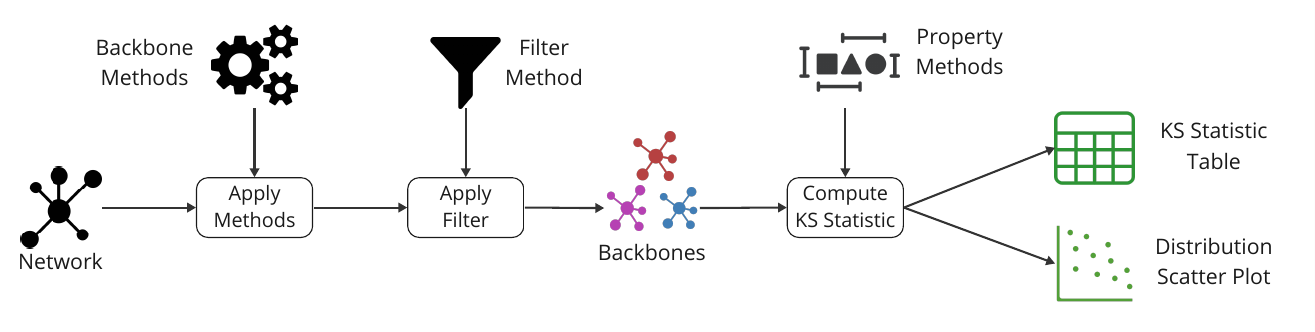}
\caption{A diagram illustrating the flow within {\fontfamily{qcr}\selectfont netbone}'s comparison framework to compute the distribution of the topological properties in the extracted backbone.} 
\label{exp3}
\end{figure}

Figure~\ref{exp3} illustrates the flow within the framework for comparing the cumulative distributions. First, we apply the global threshold method and statistical methods to the network. Then, we use the threshold filter within {\fontfamily{qcr}\selectfont netbone}'s comparison framework to filter the network. For the global threshold method, we set the threshold value to the average weight of 7000. For the statistical methods, we use a significance level of 0.05. Next, we compute the Kolmogorov-Smirnov (KS) statistic to measure the similarity between the weight and degree distributions of the original network and the backbones generated by these methods. The results are in a table for easy analysis. Additionally, one can use a distribution scatter plot to compare the distributions visually. 

Table~\ref{table:exp3} presents the KS statistic for the weight and degree distributions between the original network and the extracted backbones. The Enhanced Configuration model filter exhibits the lowest KS statistic for the weight distribution. It indicates that the backbone generated by this method closely resembles the original weight distribution, effectively highlighting the hub and spoke network topology. 
\begin{table}[ht]
        \begin{center}
            \def\arraystretch{1.2}
            \begin{tabular}{|c|c|c|}
                \hline\rule{0pt}{12pt}
                \textbf{Method} & \textbf{Weight} & \textbf{Degree} \\
                \hline
                Global Threshold Filter & 0.80& 0.40\\
                \hline
                Marginal Likelihood Filter & 0.55 & 0.41\\
                \hline
                Noise Corrected Filter & 0.51 & 0.54\\
                \hline
                Disparity Filter & 0.70 & 0.49 \\
                \hline
                Enhanced Configuration Model Filter & 0.32 & 0.55\\
                \hline
                Locally Adaptive Network Sparsification Filter & 0.66 & 0.66\\
                \hline
            \end{tabular}
            \caption{\textbf{The KS statistic comparing the weight and degree distribution between the original network and the extracted backbones was calculated using {\fontfamily{qcr}\selectfont netbone}'s comparison framework.}}
        \label{table:exp3}
        \end{center}
        \end{table}

\begin{figure}[ht]
\centering
\includegraphics[width=\linewidth]{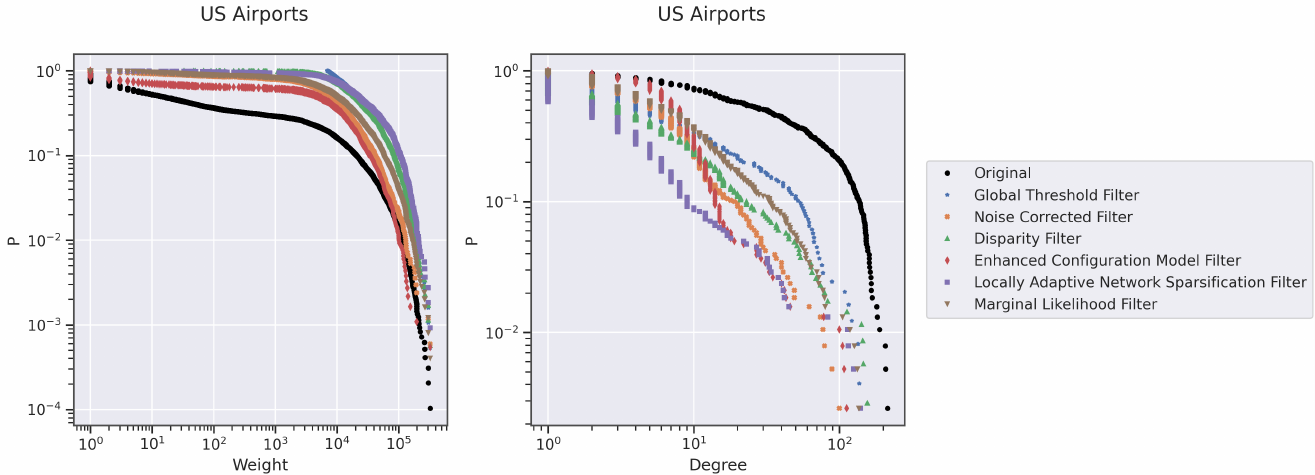}
\caption{Scatter charts that display the original network's cumulative weight and degree distribution and its extracted backbone plotted using {\fontfamily{qcr}\selectfont netbone}.} 
\label{exp3.2}
\end{figure}

On the other hand, the Marginal Likelihood filter shows the lowest KS statistic for the degree distribution. This suggests that this method better preserves the degree distribution of the original network. Users seeking to retain the nearest degree distribution can consider the Marginal Likelihood filter. One can use {\fontfamily{qcr}\selectfont netbone}'s to plot them in a scatter plot to compare the distributions visually. Figure~\ref{exp3.2} illustrates the results of this visualization, showcasing the distributions obtained from the backbone extraction methods.
This use case involves filtering the transportation network to find a backbone with weight or degree distributions that closely match the original network. {\fontfamily{qcr}\selectfont netbone}'s comparison framework is crucial in selecting the most suitable backbone extraction method. Through this framework, we compare the distributions of different methods and identify the Enhance Configuration Model filter as the closest match for the weight distribution and the Marginal Likelihood filter as the closest match for the degree distribution. 

\subsection*{Experiment 4}
The statistical methods used for backbone extraction in {\fontfamily{qcr}\selectfont netbone} are based on different null models, each aiming to understand the distribution or generation of weights in the network. As a result, these null models yield different backbones. {\fontfamily{qcr}\selectfont netbone} allows us to compute the intersection of various Backbone extraction methods. Extracting common nodes and edges across all the methods allows for observing a “consensual backbone”.

\begin{figure}[ht]
\centering
\includegraphics[width=\linewidth]{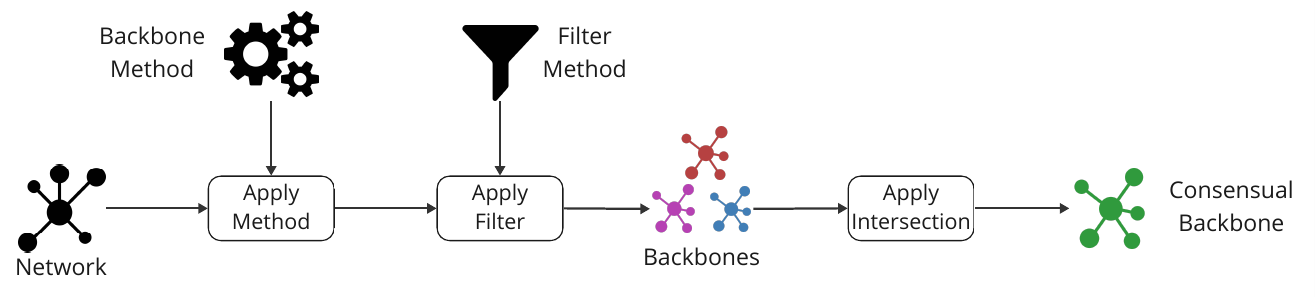}
\caption{A diagram illustrating the flow within {\fontfamily{qcr}\selectfont netbone}'s comparison framework to extract the consensual backbone from the extracted backbones.} 
\label{exp4}
\end{figure}

\begin{figure}[ht]
\centering
\includegraphics[width=\linewidth]{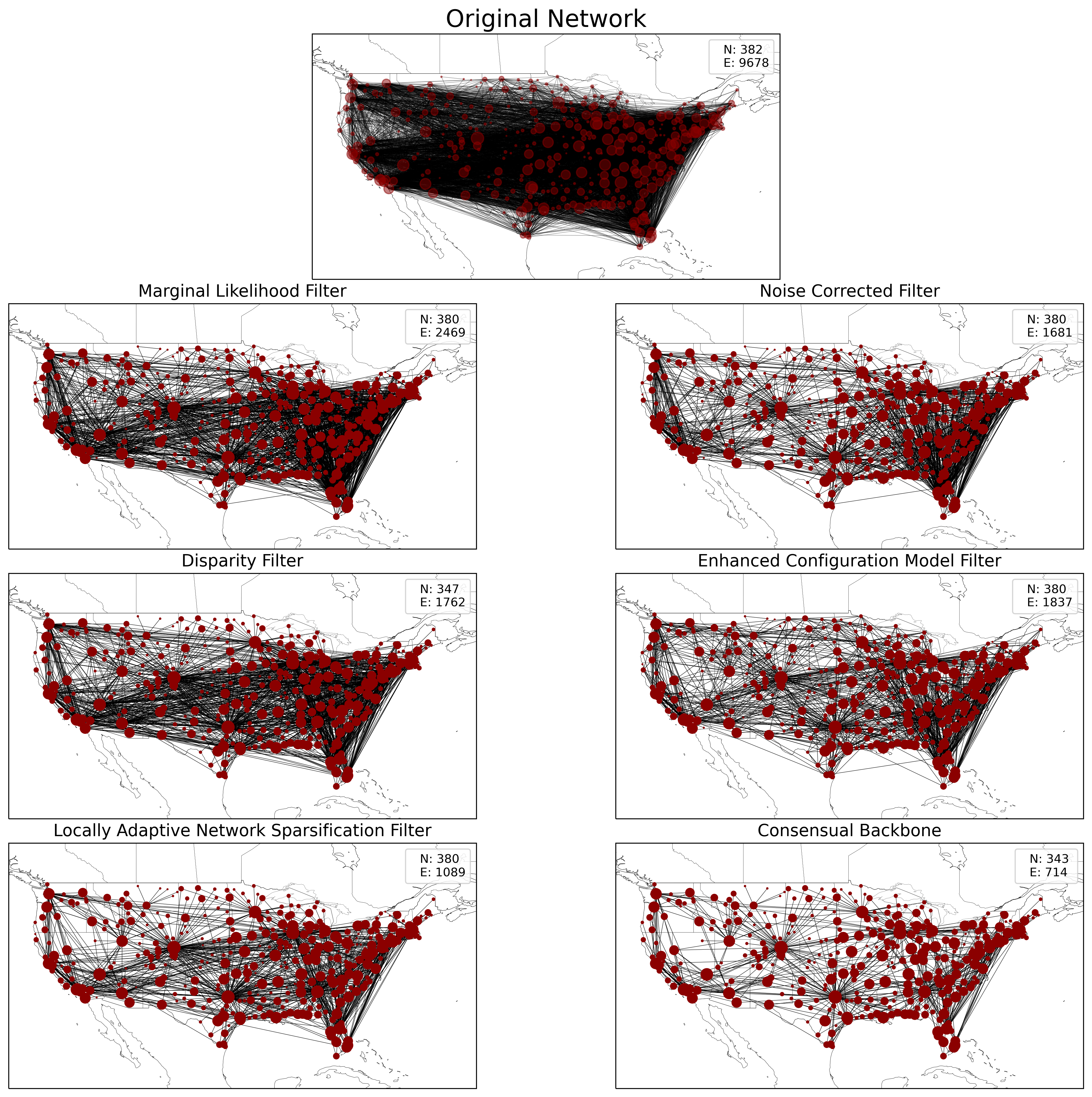}
\caption{The US air transportation network and its corresponding backbones, extracted using {\fontfamily{qcr}\selectfont netbone}. N and E are the number of nodes and edges, respectively.} 
\label{exp4.2}
\end{figure}

In this experiment, we use {\fontfamily{qcr}\selectfont netbone}'s comparison framework to extract the consensus backbone using the statistical backbone extraction methods. The process flow, depicted in Figure~\ref{exp4}, outlines the steps in extracting the consensus backbone. Firstly, we apply the statistical methods and filter them using the threshold filter with a significance level of 0.05. Then, we extract the consensus backbone by taking the intersection of the extracted backbones.
Figure~\ref{exp4.2} provides a visual representation of the extracted backbones, showcasing the number of preserved nodes and edges in each method. The consensual backbone includes 343 nodes and 714 edges. These nodes and edges hold significant value when considering the various null models used by the statistical methods. Comparing the consensual backbone to the other techniques, we can observe that it prominently highlights the hub and spoke network structure more effectively than the individual statistical methods.
This experiment demonstrates that {\fontfamily{qcr}\selectfont netbone}'s statistical consensual backbone effectively emphasizes the hub and spoke network structure compared to individual statistical backbones. Moreover, it's worth noting that the consensual extraction method is not restricted to statistical methods. Users can also use it with structural methods, allowing for various combinations and variations of backbones to explore and analyze the distinctive characteristics of these consensual backbones. 
\FloatBarrier
\subsection*{Experiment 5}
This experiment illustrates how users can integrate their custom backbone extraction method and custom evaluation properties into {\fontfamily{qcr}\selectfont netbone}'s comparison framework. To illustrate this process, we define the {\fontfamily{qcr}\selectfont new\_backbone\_method()} function. It generates random values and keeps them in a new edge property named {\fontfamily{qcr}\selectfont new\_score}. The function should return a new instance of the {\fontfamily{qcr}\selectfont Backbone} class. To initialize an instance of the {\fontfamily{qcr}\selectfont Backbone} class, users should provide a {\fontfamily{qcr}\selectfont networkx} graph containing the new edge scores, the name of the new method, the edge property name. If the edge property name represents a p-value, it should be set to {\fontfamily{qcr}\selectfont True}. Otherwise, it should be {\fontfamily{qcr}\selectfont False}. Next, users must specify an array of compatible filters. Given that the edge property is a numerical value, the appropriate filters to use in this case are the {\fontfamily{qcr}\selectfont threshold\_filter} and the {\fontfamily{qcr}\selectfont fraction\_filter}. Lastly, the {\fontfamily{qcr}\selectfont filter\_on} parameter should indicate whether the filter is applied to edges or nodes.\\\\ 
{\fontfamily{qcr}\selectfont
> from netbone.filters import threshold\_filter, fraction\_filter\\
> from netbone.backbone import Backbone\\
> import random\\\\
> def new\_backbone\_method(graph):\\
\textcolor{white}{.....}for u,v in graph.edges():\\
\textcolor{white}{........}graph[u][v]['new\_score'] = round(random.uniform(0, 1), 2)\\
\textcolor{white}{.....}return Backbone(graph,\\\textcolor{white}{.....................}method\_name='New Backbone Method',\\\textcolor{white}{.....................}property\_name='new\_score',\\\textcolor{white}{.....................}ascending=False,\\\textcolor{white}{.....................}compatible\_filters=[threshold\_filter, fraction\_filter], \\\textcolor{white}{.....................}filter\_on='Edges')}\\\\
Once the new backbone extraction method is defined, one can easily apply the method and add it to the comparison framework using the {\fontfamily{qcr}\selectfont add\_backbone()} method.
Users can now continue by adding the evaluation {\fontfamily{qcr}\selectfont measures} from the built-in methods in {\fontfamily{qcr}\selectfont netbone} or by implementing their new custom evaluation measure. To illustrate this, we define the {\fontfamily{qcr}\selectfont new\_property\_method()} method. This method will imitate the {\fontfamily{qcr}\selectfont node\_fraction()} method; it returns the node fraction preserved in the backbone. The method should take two inputs, the original and the backbone graphs. And it should return the computed property value. \\\\
{\fontfamily{qcr}\selectfont
> def new\_property\_method(original, backbone):\\
\textcolor{white}{.......}return len(backbone)/len(original)}\\\\

\section*{Conclusion}
In conclusion, {\fontfamily{qcr}\selectfont netbone} is a powerful, free, open-source Python package. It offers a variety of statistical and structural methods for extracting network backbones. Its filters can meet all use cases, and the comparison framework is a standout feature. It enables users to compare backbones with a wide range of evaluation measures. The Five experiments conducted in this paper illustrate the wide range of possible scenarios that can be analyzed using {\fontfamily{qcr}\selectfont netbone}. The first experiment showcases how the comparison framework can assist in evaluating the backbone extraction methods by comparing various topological properties. The second experiment highlights how the framework could aid users in determining the appropriate fraction or threshold for extracting backbones. The third experiment illustrates how users evaluate the distribution of property values of the extracted backbones. The fourth experiment introduces the consensual backbone and how users can create unlimited combinations of the backbone techniques. Finally, the fifth experiment illustrates how users can integrate their backbone extraction methods and custom evaluation measures into the comparison framework. Overall, the comparison framework provides users a valuable tool for comparing backbone extraction methods. In its current developmental phase, our primary focus has been on integrating classical backbone extraction methods for unweighted and weighted networks into the package. However, many situations are well described by bipartite networks. For example, in social networks, one can connect users with events. In recommendation systems, users are linked to items. Converting these bipartite graphs into a static network by projections removes important information. Tumminello, Neal, and others~\cite{tool4, backbone10} have proposed backbone extraction techniques specifically designed for bipartite projections to address this limitation. A key objective is to extend the {\fontfamily{qcr}\selectfont netbone} package by incorporating these specialized approaches. Another crucial extension concerns temporal networks. Indeed, aggregating network snapshots into a static representation entails a substantial loss of information. It can lead to conventional classical methods of weighted backbone extraction neglecting significant aspects of the underlying network. Kobayashi and others~\cite{backbone34} have introduced backbone extraction methods for temporal networks in response to this challenge. Consequently, our second goal is to enrich further the {\fontfamily{qcr}\selectfont netbone} package by incorporating these advanced techniques. Lastly, in a third major development direction, we aim to expand the scope of {\fontfamily{qcr}\selectfont netbone} to Multilayer networks. We believe these extensions will provide a comprehensive package enabling handling a broader network analysis range.

\section*{Data and Methods}
This section introduces the data and methods used in the toy example and the three experiments to evaluate the backbone extraction methods.
\subsection*{Data}
In this subsection, we introduce the networks used in the experiments, Table~\ref{data:table} reports their basic topological features. 
\subsubsection*{Les Misérables}
In the Les Misérables Network~\cite{lesmis}, nodes represent actors in Victor Hugo’s novel. They are connected if they appear in the same chapter of the Les Misérables novel. Edge weights denote the number of such occurrences.

\subsubsection*{ US Air Transportation}
In the US Air Transportation Network~\cite{backbone2}, nodes represent airports in the continental US, and edges represent the routes between these airports. Edge weights correspond to the number of passengers for the year 2006.

\begin{table}[ht]
        \begin{center}
            \def\arraystretch{1.2}
            \begin{tabular}{|c|c|c|c|c|}
                \hline\rule{0pt}{12pt}
                \textbf{Network} & \textbf{N} & \textbf{E} & \textbf{<k>} & \textbf{$\rho$}\\
                \hline
                Les Misérables & 77 & 254 & 6.5 & 0.087\\
                \hline
                US Air Transportation & 380 & 9678 & 50.9 & 0.134\\ 
                \hline
            \end{tabular}
            \caption{\textbf{The Topological features of the Les Misérables and US Air Transportation networks. N is the number of nodes. $|E|$ is the
number of edges. $<k>$ is the average degree. $\rho$ is the density.}}
        \label{data:table}
        \end{center}
        \end{table}

\subsection*{Methods}
In this subsection, we present the evaluation measures used in the experiments to evaluate the extracted backbones.
    \subsubsection*{Node Fraction}
    The node fraction in the backbone represents the proportion of nodes retained from the original network.
    \subsubsection*{Edge Fraction}
    The edge fraction in the backbone represents the proportion of edges retained from the original network.
    \subsubsection*{Weigh Fraction}
    The weight fraction in the backbone represents the proportion of edge weights retained from the original network.
    \subsubsection*{Average Degree}
    The average degree is the sum of the degrees of all network nodes divided by the number of nodes in the network.
    \subsubsection*{Density}
    The density is the ratio between the edges present in a network and the maximum number of edges that the network can contain.
    \subsubsection*{Reachability}
     The Reachability~\cite{reachability} quantifies the connectivity between any pair of nodes in a network. It is defined as the fraction of node pairs that can communicate with each other. This reads: 
    \begin{equation}
        R = \frac{1}{n(n-1)}\sum_{i\neq j \in G}{R_{ij}}.
    \end{equation}
    with $n$ is the number of nodes and $R_{ij} = 1$ if path exists between node $i$ and $j$ and $R_{ij}=0$ otherwise. The Reachability values are in the $[0, 1]$ range. If any pair of nodes can communicate in a network, the reachability $R$ becomes $1$. If $R=0$ it means all nodes are isolated from each other. 
    \subsubsection*{Two-Sample Kolmogorov-Smirnov}
    The two-sample Kolmogorov-Smirnov test (KS test)~\cite{ks:statistic} allows testing whether two samples follow the same distribution. Simply put, the KS statistic for the 2-sample test is the greatest distance between each sample's CDFs (Cumulative Distribution Function). Thus, the Kolmogorov-Smirnov statistic $D$ is given by:
    \begin{equation}
        D_{m,n} = \max_{x}|F(x)-G(x)|
    \end{equation}
    where $F(x)$ and $G(x)$ represent the CDF of the two samples, and $n$ and $m$ are the numbers of observations of the first and second samples, respectively.


\section*{Data Availability}
All data used in the toy example and the experiments are available at\\ \url{https://gitlab.liris.cnrs.fr/coregraphie/netbone/tree/main/examples/data}.

\section*{Code Availability}
NetBone is distributed via the pypi package index (\url{https://pypi.org/project/netbone/}) and is developed publically on GitLab (\url{https://gitlab.liris.cnrs.fr/coregraphie/netbone}). The examples section in the repository contains the code required to reproduce the results presented in this manuscript.

\bibliography{mybib}

\section*{Acknowledgements}
This material is based upon work supported by the Agence Nationale de Recherche under grant ANR-20-CE23-0002.

\section*{Author contributions statement}
A.Y. and A.H. designed, implemented and tested NetBone. A.Y. conducted the experiments, analyzed the results, and prepared all figures and tables. All authors participated in the formulation and writing of this paper. All authors approved the final manuscript.


\section*{Additional information}
\textbf{Competing interests}: 
All authors declare that they have no conflicts of interest.



\end{document}